\documentclass{aa}
\usepackage{graphicx}
\usepackage{natbib}
\usepackage{color}
\newcommand{\tw}{\color{black}}
\usepackage{txfonts}
\begin{document}
   \title{An optimization principle for the computation of MHD equilibria in the solar corona}

   \subtitle{}

   \author{T. Wiegelmann
          \inst{1}
          \and
          T. Neukirch \inst{2}
          \fnmsep\thanks{}
          }

   \offprints{T. Wiegelmann}

   \institute{Max-Planck-Institut f\"ur Sonnensystemforschung,
Max-Planck-Strasse 2, 37191 Katlenburg-Lindau, Germany\\
              \email{wiegelmann@mps.mpg.de}
     \and
       School of Mathematics and Statistics,
   University of St. Andrews,
   St. Andrews, KY16 9SS,
   United Kingdom \\
 \thanks{}
             }

   \date{A\&A 457, 1053-1058 (2006)}


  \abstract
   {We develop an optimization principle for computing
   stationary MHD equilibria.}
  {
 Our code for the self-consistent computation of the coronal magnetic
 fields and the coronal plasma uses non-force-free MHD equilibria.
 Previous versions of the code have been used to
 compute non-linear force-free coronal magnetic fields from photospheric measurements.
 The program uses photospheric vector magnetograms and coronal EUV images as input.
 We tested our reconstruction code with the help of a semi-analytic MHD-equilibrium.
 The quality of the reconstruction was judged by comparing the exact and reconstructed solution
 qualitatively by  magnetic field-line plots and EUV-images and quantitatively
 by several different numerical criteria.
}
   {
   Our code is able to reconstruct the semi-analytic test equilibrium with high
   accuracy. The stationary MHD optimization code developed here has about the
   same accuracy as its predecessor, a non-linear force-free optimization code.
   The computing time for MHD-equilibria is, however, longer than for force-free
   magnetic fields. We also extended a well-known class of nonlinear force-free equilibria
   to the non-force-free regime for purposes of testing the code.
   }
   {
   We demonstrate that the code works in principle using tests with analytical equilibria, but it still needs to be applied to
   real data.
    }

 \authorrunning{Wiegelmann et al.}
 \titlerunning{Optimization code}
   \maketitle
%

\section{Introduction}
Understanding many physical phenomena in the solar corona requires detailed
knowledge of the properties of the coronal magnetic field and plasma. Usually
neither the coronal magnetic field nor the plasma density, pressure, temperature, and
flow speed are directly observable. Direct observations of the coronal magnetic
field are difficult because the high temperature broadens the line-profile orders of
magnitude above the Zeeman splitting. The {\tw optically-thin} coronal line emission
has a line-of-sight integrated character, which complicates the computation of any
plasma quantities. A way out of this dilemma is the use of coronal models, which are
fed with as much observations as possible. The model assumptions, e.g. an MHD model,
constrains the non-observed quantities. Often it is not the full set of MHD equations
used, but a subset.

The simplest approach for coronal magnetic field calculations is
to assume a potential magnetic field \citep[e.g.][]{schmidt64,semel67}. The
only observational input required is the line-of-sight photospheric magnetic field,
such as observed by SOHO/MDI. These source surface potential field models provide a
first impression regarding the global coronal magnetic field, e.g. regarding the
location of coronal holes and active regions.  Details of the magnetic field
structure are often not well-approximated by potential field models, {\tw
particularly in} active regions. Approaches using force-free magnetic fields (with
electric currents parallel to the magnetic field) show results which are
significantly better. The low plasma beta in the corona $(\beta \approx 10^{-4})$
justifies that approach in the low and middle corona, but not in the photosphere
$(\beta \approx 1)$ where non magnetic forces like pressure gradients become
important. A popular simplification of force-free fields are linear force-free
fields \citep[e.g.][]{chiu:etal77,seehafer78}) where the electric current flow is
parallel to the magnetic field with a global constant of proportionality $\alpha$.

A more sophisticated approach is to allow $\alpha$ to change in space, the so-called
nonlinear force-free approach. The calculation of non-linear force-free fields
\citep[e.g.][]{sakurai81,wu:etal90,wheatland:etal00,yan:etal00,regnier:etal02,wiegelmann:etal03,wiegelmann04,
wheatland04,valori:etal05,amari:etal06,wiegelmann:etal06,inhester:etal06,schrijver:etal06}
is particularly challenging due to the intrinsic nonlinearity of the underlying mathematical problem.
From an observational point of view the nonlinear reconstruction is also more
demanding because photospheric vector magnetograph data are required to determine
the boundary conditions.

 A comparison of the magnetic loops measured by \citet{solanki:etal03} and \citet{lagg:etal04} with
different extrapolated field models by \citet{wiegelmann:etal05} revealed that linear
force-free fields are better than potential fields, but non-linear force-free models
are even more accurate.
{\tw Usually the coronal magnetic field cannot be measured directly, also some
 progress has been made recently by using magnetically sensitive coronal line emission,
 e.g., by \citet{lin:etal04} who did spectropolarimetric measurements of the
 forbidden Fe XIII 1075 nm infrared coronal line. The influence of the coronal
 magnetic field onto the emitting plasma is however observed routinely.}

Images of the coronal plasma emission are obtained by the radiation in different
wave lengths, e.g. in EUV for SOHO/EIT and TRACE. The emission is obtained from
various elements, e.g. Fe XII or Fe IX, with different instrument channels sensitive
to emission generated at different plasma temperatures. The corresponding images
have a line-of-sight character because of the corona is optically  thin. A rather
good approximation of the coronal temperature is sometimes possible, because the
radiation only occurs in a specific temperature range. Doppler shifts in the line
profile observed with e.g. SOHO/SUMER provide insight into the plasma flow speed in
the line-of-sight direction. The {\tw  high electrical conductivity} of the coronal
plasma ensures that the plasma is frozen into the magnetic field. This basically
allows us to see (the effects of) magnetic field lines in EUV-images and even use
the visible plasma radiation to improve coronal magnetic field models, e.g. to
specify the optimal value $\alpha$ for linear force-free models \citep[see
e.g.][]{wiegelmann:etal02,carcedo:etal03,marsch:etal04,wiegelmann:etal05b}.

For the low $\beta$ corona, it is also helpful that the back reaction of the plasma
onto the magnetic field can be neglected.  \cite{marsch:etal04} used a linear
force-free coronal magnetic field model and Doppler maps from SOHO/SUMER to
investigate the plasma flow in active regions. \cite{wiegelmann:etal05a} and
\cite{tu:etal05a,tu:etal05} used SUMER Doppler maps with potential and linear
force-free models to study the outflow of the solar wind in coronal holes. The next
natural step is to use more sophisticated magnetic field models, where the magnetic
field and plasma to compute self-consistently in one model, as proposed here.
In this paper we first present the basic equations used in Sect. \ref{equations}
(supplemented by a couple of appendices). We then present the algorithm based on the
equations (Sect. \ref{algorithm}) and the derivation of the non-force-free MHD
equilibria we used for testing the code (Sect. \ref{equilibrium}). The results are
presented in Sect. \ref{results}, followed by a discussion and conclusions in Sect.
\ref{discussion}.
\section[]{Basic equations}
\label{equations}
The  MHD equilibrium equations (here without gravity for simplicity
 \footnote{{\tw While it is,
     in principle, possible to include gravity in the optimization principle,
     it is hard to find (semi)-analytic equilibria to test the code. For
     configurations that are small compared to the gravity scale height of some
     $0.1$ solar radii, gravity might be neglected in first order, however.
     A consistent treatment of large-scale
     (some solar radii) structures, like helmet streamers, require not
     only the inclusion of plasma pressure and gravity, but also the
     use of spherical geometry, see e.g., \cite{wiegelmann:etal98}, but such
     computations are well beyond the scope of this paper.}}) are
\begin{eqnarray}
(\nabla \times {\bf B}) \times {\bf
B}
-\mu_0\nabla p &=& {\mathbf{0}}  \label{1} \\
\nabla \cdot {\bf B} &=& 0,  \label{2} \\
\end{eqnarray}
We have
$$
{\bf B} \cdot \nabla p=0,
$$
i.e. the pressure is constant along magnetic field lines. We show in Appendix
\ref{appendixb} that it is possible in principle  to include field-aligned
incompressible flows into the method. We take this possibility into account by
replacing the plasma pressure $p$ by a generalized pressure $\Pi=p+\rho v^2/2$ from
now on. We did not, however, calculate any example cases with field-aligned flow,
because one would need more information to disentangle the contributions of the
plasma pressure and the energy density of the flow.

The general form of the MHD equilibrium equations is given by
\begin{eqnarray}
(\nabla \times {\bf B}) \times {\bf B} &=&\nabla \Lambda  \label{L1} \\
\nabla \cdot\mathbf{B} =0, \label{L2}
\end{eqnarray}
with the special cases
\begin{equation}
\Lambda = \; \left \{
\begin{array}{l@{\qquad}l}
p_0=\mbox{ const.} & (\mbox{force-free equilibria}) \label{def_ff}  \\
\nabla (\mu_0 \, p) & (\mbox{MHS equilibria}) \label{def_mhs} \\
\nabla \left(\displaystyle\frac{\mu_0 \, \Pi}{1-M_A^2} \right) & (\mbox{field-aligned flow, constant $M_A$}) \label{def_mhd}
\end{array}
\right. . \label{L1a}
\end{equation}
To solve  Eqs. (\ref{L1}) and (\ref{L2}), we define the functional
\begin{equation}
L({\bf B},\Lambda)=\int \frac{w_a}{B^2} \left|(\nabla \times {\bf B}) \times {\bf
B}- \nabla \Lambda \right|^2 + w_b |\nabla \cdot {\bf B}|^2 \; d^3x,
\label{defL_lambda}
\end{equation}
where $w_a$ and $w_b$ are positive definite weighting functions
 \footnote{The
functions $w_a$ and $w_b$ can e.g. be used to deal with unknown boundary conditions
\citep[see][]{wiegelmann04}. To test the method, it is sufficient to use
$w_a=w_b=1$.}.

It is obvious that
Eqs. (\ref{L1}) and (\ref{L2}) are satisfied if the functional (\ref{defL_lambda}) reaches its
minimum at $L=0$. The functional (\ref{defL_lambda}) generalizes the force-free approach of
\citet{wheatland:etal00} and the magnetohydrostatic model of
\citet{wiegelmann:etal03a}.

To obtain evolution equations for the magnetic field and the generalized plasma pressure, we take the derivative of Eq. (\ref{defL_lambda}) with respect to an artificial
 parameter $t$, assuming that both ${\bf B}$ and $\Lambda$ depend on $t$:
\begin{eqnarray}
\frac{1}{2} \; \frac{d L}{d t} &=& -\int_{V} \frac{\partial {\bf B}}{\partial t}
\cdot {\bf \tilde{F}}
-\nabla \cdot {\bf \Omega_a} \, \frac{\partial \Lambda}{\partial t} \; d^3x \nonumber \\
&&-\int_{S} \frac{\partial {\bf B}}{\partial t} \cdot {\bf \tilde{G}} + {\bf
\Omega_a} \cdot {\bf \hat n} \, \frac{\partial \Lambda}{\partial t} \; d^2x ,
\label{minimize1a}
\end{eqnarray}
 {\tw where ${\bf \tilde{F}}$ and ${\bf \tilde{G}}$ are defined in Appendix \ref{appendixa}.}

 If ${\bf B}$ and $\Lambda$ are kept fixed on the boundary of the computational
box, the surface integral vanishes and we can minimize $L$ by the solving the
equations
\begin{eqnarray}
\frac{\partial {\bf B}}{\partial t} &=& \mu \, {\bf \tilde{F}} \label{iterateB1} \\
\frac{\partial \Lambda}{\partial t} &=& -\nu \, \nabla \cdot {\bf \Omega_a}
\label{iterateLambda1}
\end{eqnarray}
iteratively, with positive constants $\mu$ and $\nu$
(see Appendix \ref{appendixa} for the mathematical derivation).
The form of Eqs.
(\ref{iterateB1}) and (\ref{iterateLambda1}) ensures that $L$ decreases
monotonically during the
computation of ${\bf B}$ and $\Lambda$.
\section{Algorithm}
\label{algorithm} To compute nonlinear, selfconsistent 3D-MHD equilibria,  we use
the following steps:
\begin{enumerate}
\item Compute a potential field. This can be done from the $B_z$ component
of the vector magnetogram alone.
\item Distribute the plasma (or, say, the generalized plasma pressure $\Lambda$)
along the potential magnetic field by solving
\begin{equation}
{\bf B} \cdot \nabla \Lambda =0
\label{Bgradlambda}
\end{equation}
 with an upwind method (used here) or a magnetic field line tracer
 \footnote{By multiplying Eq. (\ref{def_ff}) with ${\bf B}$, we get
 ${\bf B} \cdot \nabla \Lambda =0$, which implies that the generalized pressure
 $\Lambda$ is constant on magnetic field lines. This kind of equation ${\bf B}
\cdot \nabla \alpha =0$ is also used in Grad-Rubin like extrapolation codes to
distribute the force-free parameter $\alpha$ along the field lines in space.}.

\item Substitute the boundary values of the computational box from the observed
vector magnetogram. The interior of the computational box remains filled
with a potential field and a corresponding plasma distribution.
\item Iterate for ${\bf B}$ and $\Lambda$ by Eqs. (\ref{iterateB1}) and
(\ref{iterateLambda1}). The continuous form of these equations ensures that $L$ is
monotonically decreasing. This is also ensured in the discretized form if the
iteration step $dt$ is small enough. The code automatically controls the optimal
iteration step. If $L(t+dt) \geq L(t)$, the step is refused and repeated with $dt$
reduced by a factor of two. We increase $dt$ by a factor $1.01$ after each
successful iteration step to allow $dt$ to be as large as possible with respect to
the stability condition.
\item The iteration stops when $L$ reaches its minimum. In practise we stop the
iteration when $\frac{\partial L}{\partial t}/L < 10^{-6}$ for $100$ consecutive
steps.
\item As result we get the magnetic field ${\bf B}$ and generalized plasma pressure
$\Lambda$, which fulfill the MHD equations and is consistent with the observed
boundary conditions. In the final step one has to disentangle the generalized plasma
pressure $\Lambda$ with respect to the plasma pressure $p$ and the flow velocity
$v$. For MHS equilibria, this is trivial: $p=\frac{\Lambda}{\mu_0}$; for equilibria
with flow one needs further observations/assumptions regarding the plasma flow, e.g.
from SOHO/SUMER.
\end{enumerate}
\section{Semi-analytic test equilibrium}
\label{equilibrium}
\begin{figure*}
\mbox{
\includegraphics[height=7cm,width=9cm]{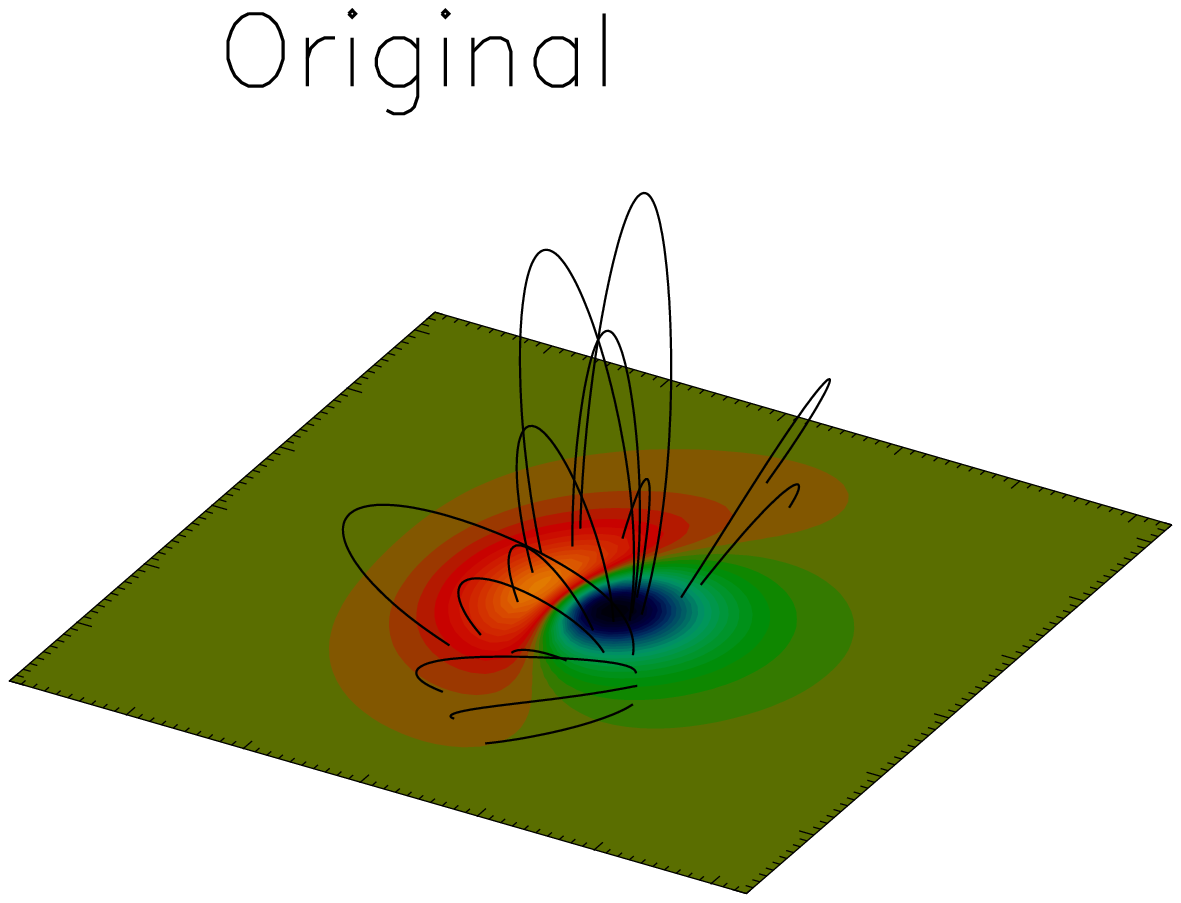}
\includegraphics[height=7cm,width=9cm]{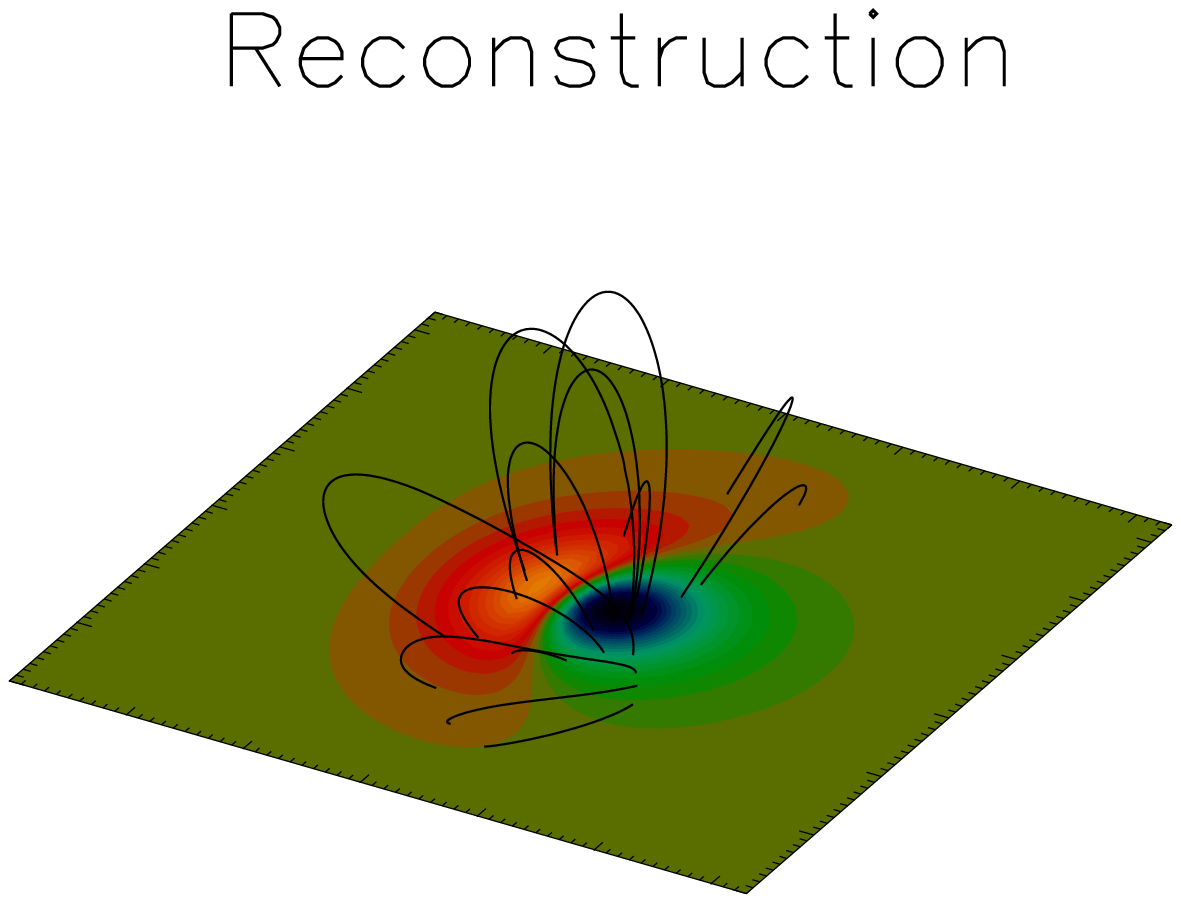}}
\caption{Left panel: original MHD-equilibrium; right panel: reconstruction. The
colour coding shows the line-of sight magnetic field on the photosphere.}
\label{fig1}
\end{figure*}
\begin{figure*}
\mbox{
\includegraphics[height=7cm,width=9cm]{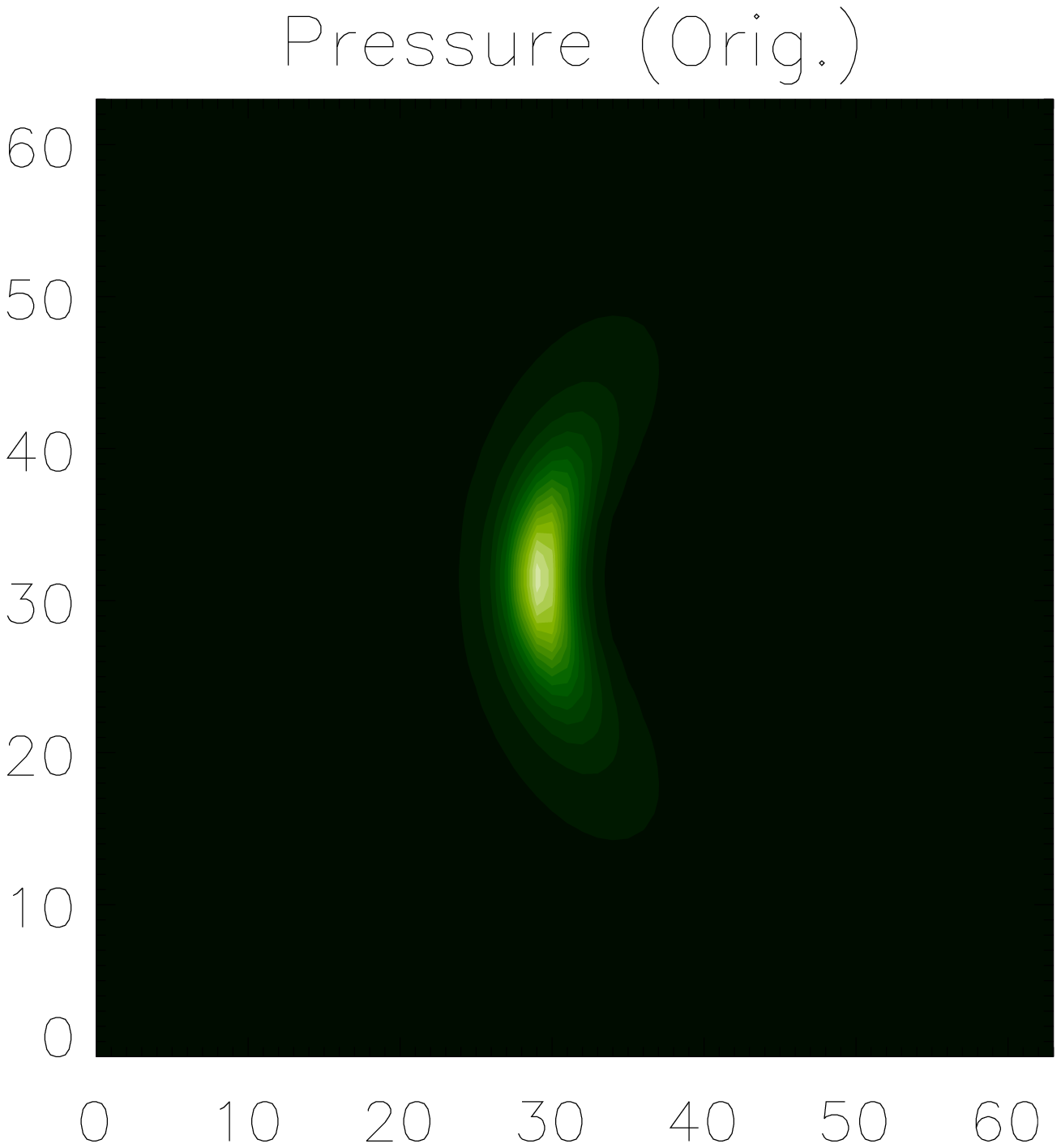}
\includegraphics[height=7cm,width=9cm]{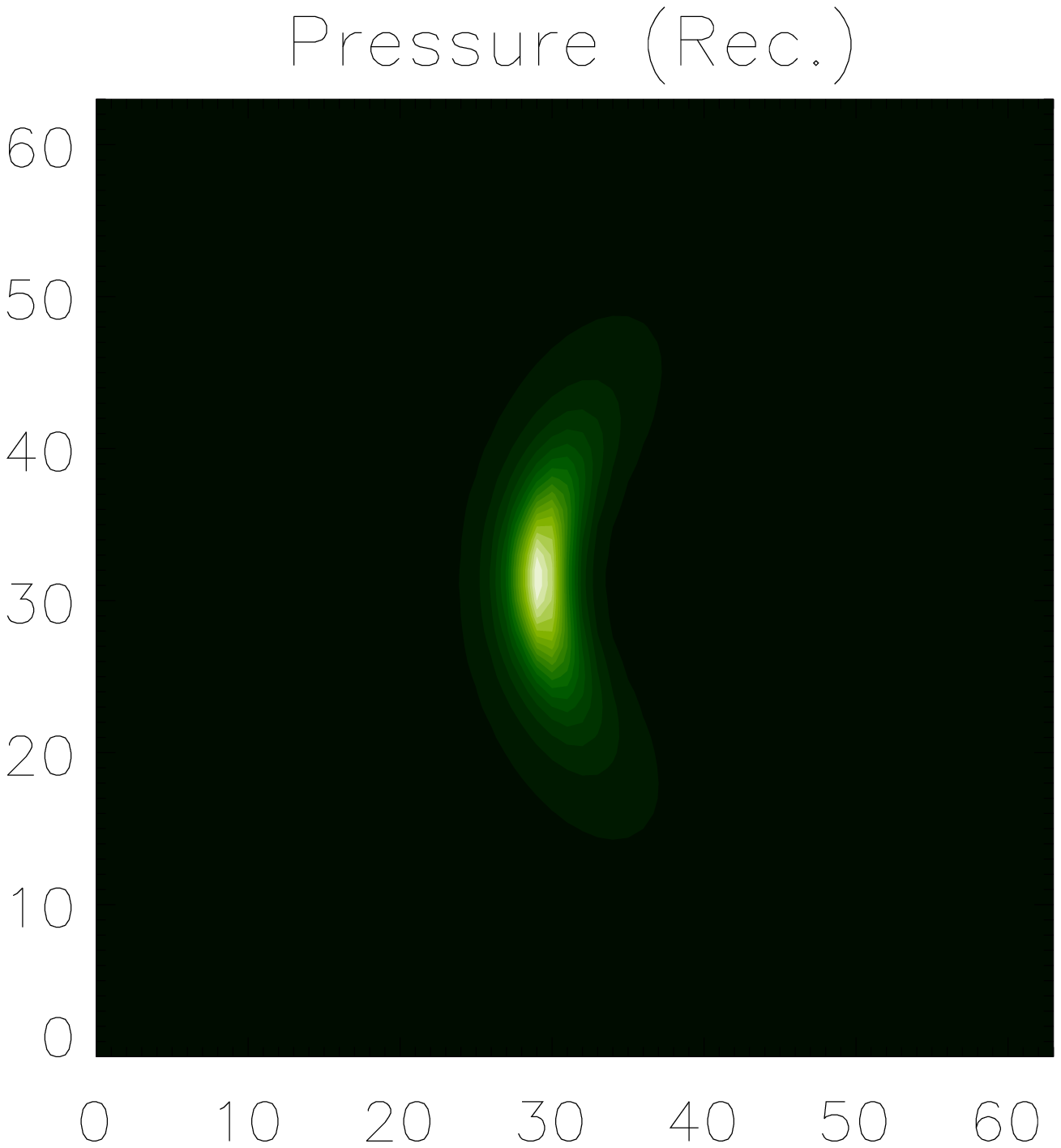}}
\mbox{
\includegraphics[height=7cm,width=9cm]{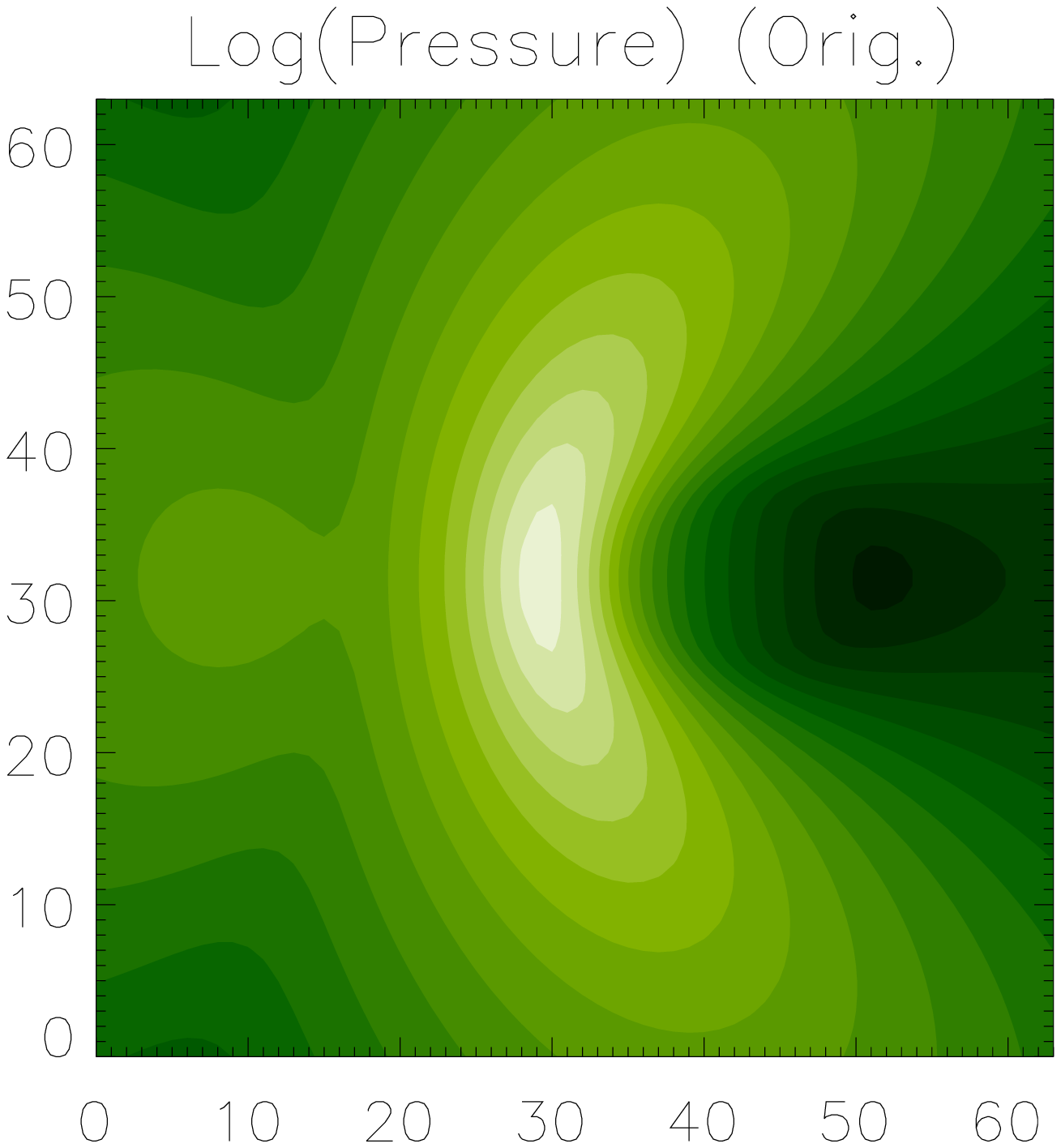}
\includegraphics[height=7cm,width=9cm]{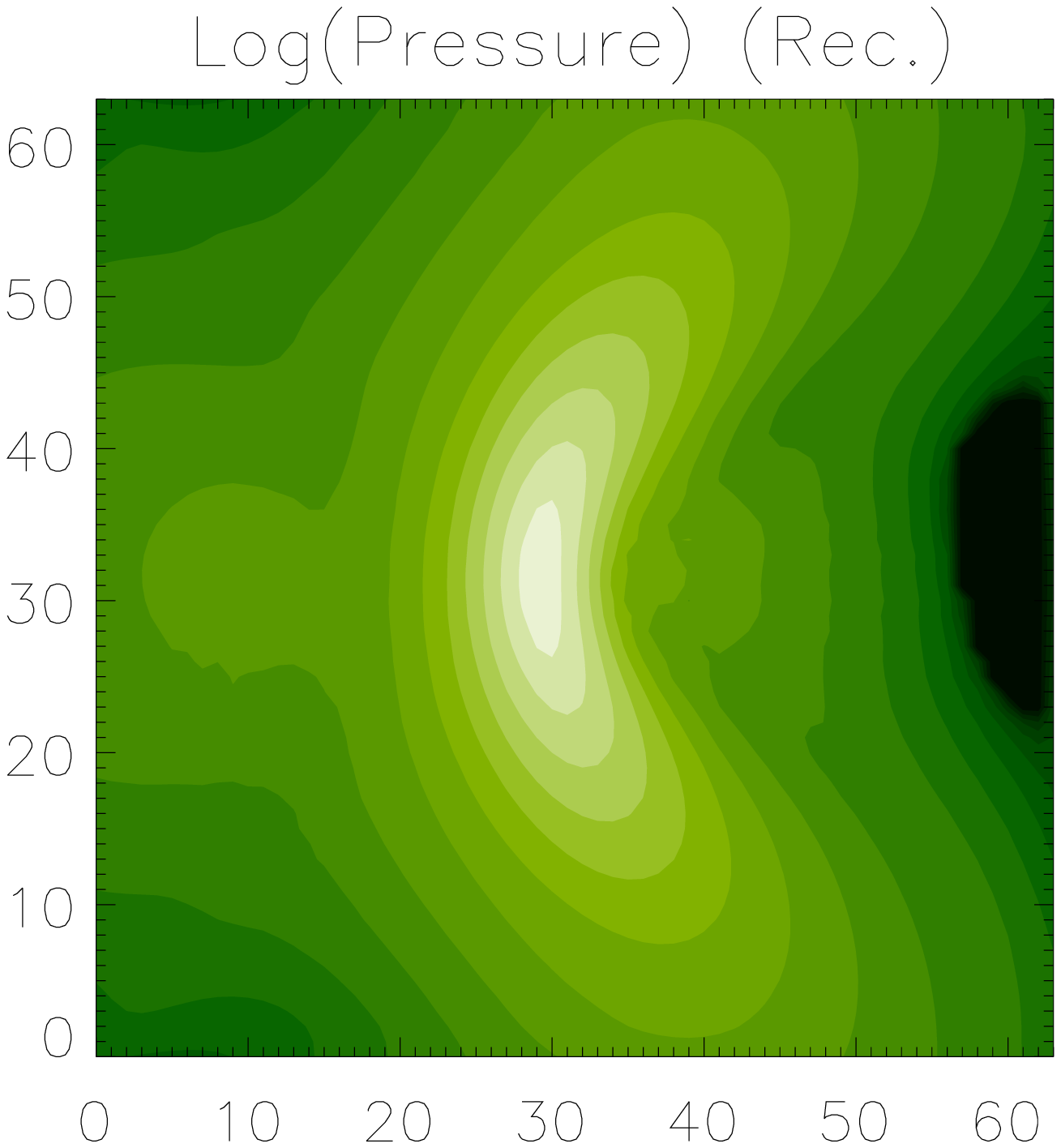}}
\caption{We show the
line-of-sight integration of $\Lambda$ along the z-axis
(Top: linear scaling, bottom logarithmic scaling). This quantity seems to be
important because coronal images have a line-of-sight integrated character as a
consequence of the optical thin coronal line emission. The left hand panels
correspond to the original solution and the right hand panels to our reconstruction.}
\label{fig2}
\end{figure*}
We test our method with the help of a semi analytic equilibrium similar to the
force-free \cite{low:etal90} solution {\tw (henceforth LL)}. We solve the
Grad-Shafranov equation for axis-symmetric force-free fields in spherical
coordinates $r$, $\theta$, $\phi$:
\begin{equation}
-\frac{\partial^2 A}{\partial r^2} -\frac{1-\mu^2}{r^n}\frac{\partial^2 A}{\partial
\mu^2} = \mu_0 r^2(1-\mu^2)\frac{d \Lambda}{dA} + b_\phi\frac{d b_\phi}{d A} .
\label{GSgeneral}
\end{equation}
LL looked for self-similar (in $r$) solutions by choosing $b_\phi(a) = c A^q$. We
generalize this approach by also choosing $\Lambda(A)=k A^s$, where the power $s$ has to
be chosen such that self-similar solutions are possible.

Following LL we assume
\begin{equation}
A(r,\mu) = \frac{P(\mu)}{r^n}. \label{llsol}
\end{equation}
 Substituting this into Eq. (\ref{GSgeneral}) and using the above expressions for $\Lambda$ and $b_\phi$, we get
 \begin{eqnarray}
- n(n+1)\frac{P}{r^{n+2}}-\frac{1-\mu^2}{r^{n+2}}\frac{d^2P}{d \mu^2} & = & \nonumber \\
& & \mbox{\hspace{-3cm}}  \mu_0 s k
(1-\mu^2) \frac{P^{s-1}}{r^{n(s-1)-2}} + q c \frac{P^{2q-1}}{r^{n(2q-1)}}.
\label{GSspecial}
\end{eqnarray}

We now determine $q$ and $s$ such that all powers of $r$ are equal, obtaining the
equations
\begin{eqnarray}
n+2 & = & n(s-1)-2 \label{teq}\\
n+2 & = & n(2q-1). \label{qeq}
\end{eqnarray}
Solving Eq. (\ref{teq}) gives
\begin{equation}
\label{tofn} s = 2 +\frac{4}{n},
\end{equation}
whereas Eq. (\ref{qeq}) gives the same result as LL
\begin{equation}
\label{qofn} q = 1 + \frac{1}{n} .
\end{equation}
The equation for $P(\mu)$ is then given by
\begin{eqnarray}
(1-\mu^2)\frac{d^2P}{d \mu^2} +n(n+1)P +
2 k \left(1 + \frac{2}{n}\right) (1-\mu^2)P^{1 +4/n} & & \nonumber \\
& & \mbox{\hspace{-4cm}}
+ a^2 \left(1 + \frac{1}{n}\right) P^{1+2/n} = 0 . \label{peq}
\end{eqnarray}
This equation is nonlinear and has to be solved numerically.

For our test equilibrium, we use a similar parameter set to the one in
 LL for the force-free case
 ($\Phi=\frac{\pi}{4}, \; l=0.3$). In Eq. (\ref{peq}) we choose (as LL)
 $a^2=0.425$. Both $\Phi$ and $l$ have the same meaning as in LL here.

 The difference of our solution to LL is, that we have
  the additional term  $2 k \left(1 + \frac{2}{n}\right) (1-\mu^2)
 P^{1 +4/n}$ which corresponds to the (generalised) plasma pressure.
 ($k=0$ corresponds to LL.) For the present paper we choose $k=10$ and
 $\frac{\partial p}{\partial \mu}=2.097$ for $\mu=-1$.

\section{Results}
\label{results}
In Fig. \ref{fig1} we compare some magnetic field lines of the original solution
(left panel) with the result of our reconstruction (right panel). The reconstructed
solution obviously agrees well with the original, and an inspection by eye only
shows hardly visible differences for some loops. Figure \ref{fig2} contains images
produced from the original (left panels) and reconstructed (right panels) plasma
pressure. The images have been produced by a line-of-sight integration along the
$z$-axis to mimic the optically thin coronal plasma. The top panels use a linear
scaling and the bottom panels a logarithmic one. While the overall structure of the
plasma emission agrees in the original and the reconstruction, there are small
deviations (visible in the logarithmic images) in darker (weak magnetic field)
regions, in particular at the right side of the image.
\begin{figure}
\includegraphics[height=7cm,width=9cm]{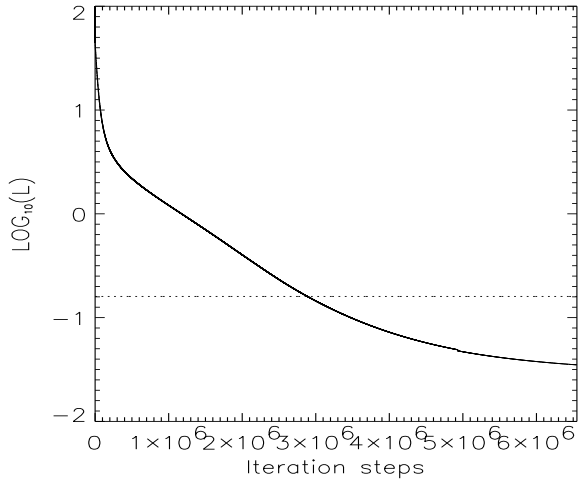}
\caption{Evolution of the functional $L({\bf B},N)$ during the iteration for a box
of $64^3$. The dotted line corresponds to the discretisation error of the analytic
solution $[L=0.16$ or $Log_{10}(L)=-0.80]$. }
\label{fig3}
\end{figure}
%
To evaluate the quality of reconstruction we used several figures of merit. The
functional $L$ as defined in Eq. (\ref{defL_lambda}) provides automatically
 a measure of how well the MHD equilibria and the solenoidal condition are satisfied.

The evolution of $L$ during the optimization process is shown in Fig. \ref{fig3} as
a function of the iteration steps. For the $64^3$ box, $L$ has decreased below the
discretization error of the analytic solution during the iteration. The number of
iteration steps (and thus the the computing time) for convergence is, however, about
1000 times greater than for a corresponding force-free optimization. The reason for
this behavior is because the semi-analytic solution has a huge difference in
magnitude of the pressure throughout the computational box with
 $p_{\rm max}=1.9133 \cdot 10^{4}$ and  $p_{\rm min}=1.0239 \cdot 10^{-21}$. The code obviously has
problems dealing with the pressure differences during the iteration. Test runs with
prescribed pressure profile (only iterating for {\bf B}) converge at the same rate
as a corresponding force-free calculation \citep[see e.g,][where the magnetic
pressure $\frac{B^2}{2}$ varies between $4.616 \cdot 10^4$ and $2.87 \cdot
10^{-2}$.]{schrijver:etal06}

\citet{schrijver:etal06} developed several figures of merit to quantify the
difference between two vector fields ${\bf B}$ (semi-analytic test field) and ${\bf
b}$ (reconstructed fields). The figures have been used to evaluate the quality of
six different non-linear force-free extrapolation codes, by comparing the result
with the LL solution. We use the same definitions as given in  section 4 of
\citet{schrijver:etal06} for evaluating the quality of the non force-free magnetic
fields here:
\begin{itemize}
\item Vector correlation
\begin{equation}
C_{\rm vec}=  \sum_i {\bf B_i} \cdot {\bf b_i}/ \left( \sum_i |{\bf B_i}|^2 \sum_i
|{\bf b_i}|^2 \right)^{1/2};
\end{equation}
\item Cauchy-Schwarz inequality
\begin{equation}
C_{\rm CS} = \frac{1}{N} \sum_i \frac{{\bf B_i} \cdot {\bf b_i}} {|{\bf B_i}||{\bf
b_i}|},
\end{equation}
where $N$ is the number of vectors in the field;
\item Normalized vector error
\begin{equation}
E_{\rm N} = \sum_i |{\bf b_i}-{\bf B_i}|/ \sum_i |{\bf B_i}|;
\end{equation}
\item Mean vector error
\begin{equation}
E_{\rm M} = \frac{1}{N} \sum_i \frac{|{\bf b_i}-{\bf B_i}|}{|{\bf B_i}|};
\end{equation}
\item Magnetic energy of the reconstructed field
normalized with the energy of the input field
\begin{equation}
\epsilon = \frac{\sum_i |{\bf b_i}|^2}{\sum_i |{\bf B_i}|^2}.
\end{equation}
\end{itemize}
The two vector fields agree perfectly if the figures of merit
($C_{\rm vec},C_{\rm CS},\epsilon$) are unity
and if ($E_N,E_M$) are zero.

We also compare how closely the generalized plasma pressure $\Lambda$ of our
reconstruction agrees with the original. To do so we compute the linear Pearson
correlation coefficient, both for the 3D plasma pressure $Corr \Lambda_{3D}$ and the
line-of-sight integration (with respect to the $z$ axis) $Corr \Lambda_{2D}$. The
latter value is a useful quantity because observed EUV-images (e.g. SOHO/EIT, TRACE)
have a line-of sight integrated character as a consequence of the optically thin
coronal plasma.

\begin{table}
\caption{Quality of the reconstruction.}
\label{table1}
\begin{tabular}{lcc}
 & Exact & Rec.  \\
 \hline
 $L$ & $0.16$ & $0.035$ \\
 Vector Correlation & $1$ & $0.999$ \\
 Cauchy Schwarz & $1$ & $0.992$ \\
 Normalized Vector Error & $0$ & $0.130$ \\
 Mean Vector Error & $0$ & $0.073$ \\
 Relative Magnetic Energy $\epsilon$ & $1$ & $0.996$ \\
 $Corr \Lambda_{3D}$ & $1$ & $0.998$ \\
 $Corr \Lambda_{2D}$ & $1$ & $0.999$ \\
\hline
\end{tabular}
\end{table}

\section{Discussion and conclusions}
\label{discussion}

We have generalized the optimization method for nonlinear force-free fields
\citep{wheatland:etal00} to magnetohydrostatic equilibria by including a pressure
gradient. Using a semi-analytical magnetohydrostatic equilibrium similar to the
nonlinear force-free equilibria by \citet{low:etal90} for testing the code,  we
showed that the optimization method also works in principle if plasma pressure is
included. The reconstructed solution agrees well with the exact solution. For
application to real data, the method will have to be
 developed further to include measured information about the plasma properties, which in the present paper
 we have taken from the exact solution. We have also shown that it is possible to include
 an incompressible field-aligned plasma flow.

Compared to the corresponding nonlinear force-free code, however, the
MHD-optimization method presented here (for ${\bf B}$ and $\Lambda$ simultaneously)
is about a factor of $10^3$ slower than force-free computation. The reason for this
seems to be the additional equation for updating the plasma pressure. If we
prescribe the correct exact pressure and solve only for the magnetic field, the
convergence speed of the method is similar to that of the corresponding force-free
case. If, on the other hand, we fix the magnetic field using the exact solution and
iterate only the pressure, the convergence speed is much less and comparable to the
convergence speed of the combined magnetic field/plasma pressure iteration. The
reason for this behavior by the method is the huge difference in the values of the
plasma pressure throughout the computational box, combined with the relatively low
values of the plasma $\beta$. This means that even small changes in the magnetic
field have to compensate for by much larger changes in the plasma pressure.

Despite these practical difficulties, we believe that the method has a lot of
potential to improve magnetic-field reconstruction by including more information
from observations.

\appendix

\section{Including field-aligned flow.}
\label{appendixb}

For completeness, we give some details here of how field-aligned incompressible
flows could in principle be included in the optimization method. Although flows seem
to be common in the corona, the inertial force scales with the square of the
Alfv\'en Mach number, which usually means that only flows with Alfv\'en Mach numbers
close to $1$ have a noticeable effect upon the equilibrium structure
\citep[e.g.][]{petrie:neukirch99}. We also notice that to use the method including
flow one would need additional information, to be able to disentangle the effects of
the plasma pressure and plasma flow. {\tw The assumption of an incompressible
field-aligned flow is an idealization, because any real flow that approaches the
base of the coronal
 loops, which has small pressure scale heights, will be very compressible,
 even when the flow is slow.}

We restrict our treatment here to incompressible field-aligned flows with constant
Alfv\'en Mach number $M_A$. For incompressible field-aligned flows, a general
transformation theory exists \citep[e.g.][]{gebhardt:etal92} that allows the mapping
of equilibria with field-aligned incompressible flow (with $M_A<1$) onto static MHD
equilibria. For further details we refer to \citet{gebhardt:etal92}.

Neglecting gravity for simplicity, the stationary incompressible MHD equations
are
\begin{eqnarray}
\rho ({\bf v}\cdot \nabla) {\bf v} &=& \frac{1}{\mu_0}(\nabla \times {\bf B}) \times {\bf
B}
-\nabla p  \label{1_app} \\
\nabla \cdot {\bf B} &=& 0  \label{2_app} \\
\nabla \cdot(\rho {\bf v}) &=& 0 \label{3_app} \\
\nabla \cdot {\bf v} &=& 0 \label{4_app} \\
\nabla \times ({\bf v} \times {\bf B}) &=& 0  \label{5_app}.
\end{eqnarray}
Equations (\ref{3_app}) and (\ref{4_app}) imply that the plasma density is constant
along magnetic field lines:
\[
{\bf B} \cdot \nabla \rho=0.
\]
Equation (\ref{5_app}) is identically satisfied for field-aligned flow ($\mathbf{v}
\parallel \mathbf{B}$, implying a vanishing electric field). The plasma velocity can
then be written as
$$
{\bf v} = M_A \; {\bf v_A}
$$
where $M_A$ is the Alfven Mach number and ${\bf v_A}$ the Alfven velocity, defined by
$$
{\bf v_A}=\frac{{\bf B}}{\sqrt{\mu_0 \rho}}.
$$
Rewriting the force balance equation with the vector identity
$$
({\bf v}\cdot \nabla ){\bf v}=\frac{1}{2} \nabla v^2 + (\nabla \times {\bf v})
\times {\bf v},
$$
it takes the form
\begin{equation}
\rho (\nabla \times {\bf v}) \times {\bf v}  = \frac{1}{\mu_0}(\nabla \times {\bf
B}) \times {\bf B}
-\nabla \left(p+\frac{\rho v^2}{2} \right)   ,
\end{equation}
and for a constant Alfven Mach number $M_A$ we immediately get
\begin{eqnarray}
(\nabla \times {\bf B}) \times {\bf B} &=&  \nabla \left(\frac{\mu_0 \Pi}{1-M_A^2} \right) \label{L1_app} \\
\nabla \cdot {\bf B} &=& 0 \label{L2_app}
\end{eqnarray}
where $\Pi=p+\frac{\rho v^2}{2}$ is the generalized pressure (plasma pressure and
dynamic pressure). Equation (\ref{L1_app}) has (at least for $M_A < 1$) a structure
that is similar to the magnetohydrostatic equilibrium equation. This is a natural
property of MHD equilibria with an incompressible stationary plasma flow, which can
be derived from MHS-equilibria by a suitable transformation as shown, for example,
in \citet{gebhardt:etal92}. Equations (\ref{L1_app}) and (\ref{L2_app}) can then, in
principle, be solved using the method presented in Sect. \ref{equations}, but more
information would be needed to obtain the plasma pressure $p$ and the density $\rho$
and velocity $\mathbf{v}$ separately.

\section[]{Mathematical details}
\label{appendixa}
 With
\begin{eqnarray}
{\bf \Omega_a} &=& \; B^{-2} \;\left[(\nabla \times {\bf B})
\times {\bf B} - \nabla \left(\frac{\mu_0 \Pi}{1-M_A^2} \right) \right] \\
{\bf \Omega_b} &=& B^{-2} \;\left[(\nabla \cdot {\bf B}) \; {\bf B} \right],
\label{defomega}
\end{eqnarray}
the functional (\ref{defL_lambda})  reads:
\begin{equation}
L=\int_{V} \; w_a \; B^2 \Omega_a^2+ w_b \; B^2 \Omega_b^2 \; d^3x . \label{defLap}
\end{equation}
We minimize equation (\ref{defLap}) with respect to an iteration parameter $t$
 and obtain an iteration equation for the magnetic field
\begin{equation}
\Rightarrow \frac{1}{2} \; \frac{d L}{d t}=-\int_{V} \frac{\partial {\bf
B}}{\partial t} \cdot {\bf \tilde{F}} \; d^3x -\int_{S} \frac{\partial {\bf
B}}{\partial t} \cdot {\bf \tilde{G}} \; d^2x \label{minimize1}
\end{equation}
\begin{eqnarray}
{\bf \tilde{F}}&=&{\bf \tilde{F_a}}+{\bf \tilde{F_b}}  \\
{\bf \tilde{G}}&=&{\bf \tilde{G_a}}+{\bf \tilde{G_b}}
\end{eqnarray}
\begin{eqnarray}
{\bf \tilde{F_a}}&=& w_a \; {\bf F_a} +({\bf \Omega_a} \times {\bf B}
)\times \nabla w_a \\
{\bf \tilde{F_b}}&=& w_b \; {\bf F_b} +  ({\bf \Omega_b} \cdot {\bf B}) \; \nabla w_b \\
{\bf \tilde{G_a}}&=& w_a \; {\bf G_a} \\
{\bf \tilde{G_b}}&=& w_b \; {\bf G_b}
\end{eqnarray}
\begin{eqnarray}
{\bf F_a} & =& \nabla \times ({\bf \Omega_a} \times {\bf B} ) - {\bf \Omega_a}
\times (\nabla \times {\bf B})
+ \Omega_a^2 \; {\bf B} \\
{\bf F_b} & =& \nabla({\bf \Omega_b} \cdot {\bf B})-  {\bf \Omega_b}(\nabla \cdot
{\bf B})
+\Omega_b^2\; {\bf B} \\
\end{eqnarray}
\begin{eqnarray}
{\bf G_a} & = & {\bf \hat n} \times ({\bf \Omega_a} \times {\bf B} ) \\
{\bf G_b} & = & -{\bf \hat n} (\bf \Omega_b \cdot \bf B) \label{defG}
\end{eqnarray}
and $\hat n$ is the inward unit vector on the surface $S$. The surface integral in
(\ref{minimize1}) vanishes if the magnetic field is described on the boundaries of a
computational box.


\label{lastpage}

\begin{acknowledgements}
 The work of Wiegelmann was supported by DLR-grant 50 OC 0501.
 TW acknowledges the warm hospitality during two research visits in
 the Solar group, university St. Andrews, UK, and financial support
 by a British Council-DAAD and a PPARC grant.
\end{acknowledgements}

\bibliographystyle{aa}
%

\end{document}